# TEACHING STATISTICS WITH EXCEL AND R

MATTEO DELL'OMODARME
*Scuola Normale Superiore di Pisa*
matt_dell@email.it

GIADA VALLE
*Department of Physics, University of Pisa*
valle@df.unipi.it

**ABSTRACT**

*Despite several deficiencies, the use of spreadsheets in statistics courses is increasingly common. In this paper we discuss many shortcomings resulting from this approach. We suggest a technique integrating a spreadsheet and a dedicated software package (R), that takes advantage of the characteristics of both programs. We also present a simple protocol of transferring data from the spreadsheet to R that can be safely used even in introductory courses.*

## 1. INTRODUCTION

Nowadays spreadsheets are broadly used for data entry and manipulation. They are common not only in the native market, i.e. business analysis, but also in such areas as biological and medical sciences. Their attractiveness is enhanced by their integration in standard packages for office automation (such as Microsoft Office and Open Office); with such software available in home and office computers, they produce a large population of potential users. Many academics are used to employing spreadsheets for treating data, and most of their students can use them, so it is understandable that such software is becoming more and more popular also as a tool for teaching statistics (see for example Ostrowski, 1988; Wood, 1992; Callender and Jackson, 1995; Hunt and Tyrrell, 1995; Hunt, 1996; Bell, 2000; Warner and Meehan, 2001), even if this approach can lead to some serious problems (for example see Hawkins, 1996; Nash and Quon, 1996).

The aim of this paper is to integrate a spreadsheet and R, a dedicated statistical package. We focus our attention mainly to Excel - the most widely used spreadsheet - but similar considerations can be extended to other spreadsheets. The structure of the paper is the following: in section 2 we briefly review some of the limitations and the conveniences of Excel, both for students and teachers; in section 3 we discuss the choice of R as a package for teaching statistics; in section 4 we present a possible way of integrating the two in a postgraduate course for students in Human Medicine and Biology; finally in section 5 we present some conclusions.

## 2. SPREADSHEETS AS A TOOL FOR TEACHING STATISTICS

Two fundamental questions arise in planning to use software in a statistics course: is it really worth using software to teach statistics and how much time will students take to master it? One common answer is that the use of software packages will allow students to enhance their knowledge on their own by doing statistical computer simulations and also gain experience appropriate to real-world environments and applications (on this and related topics see Burrill 1996). In this framework the use of spreadsheets can be interpreted as a mean to reduce the effort of introducing a totally new software package, with a huge saving of time for both teachers and students.

The widespread use of spreadsheets, the ease with which data can be entered and manipulated, the capability to import data from other applications and data sources (i.e. web-based repositories) and from ASCII text files, and the advanced filtering options are the most important reasons for their adoption. In addition, many statistical and mathematical functions are already incorporated, making this software appealing for basic statistics education. It is therefore not surprising that many introductory courses of Statistics, especially in faculties of Human and Veterinary Medicine, Psychology and Biology, rely upon Excel as the unique software tool employed during the lectures.

Due to their design, which allows immediate updating of the results when data are changed, the most important advantage in this choice is evident when teaching probability theory and distributions (see for example Ageel, 2002; Christie, 2004). Moreover graphics are refreshed after data change, allowing students to play with distribution functions, to easily identify patterns in two-dimensional data, and to discover possible outliers. An appropriate use of labelling and the ability to inspect the formulae entered in the cells give students and teachers the ability to check all the steps of the analysis without difficulty.

Some classical tests (e.g. Student t test, z test, correlation test, one and two way ANOVA with and without replications, regression), can be covered with the help of several statistical add-in macros supplied with Excel. Their pedagogical use is somewhat controversial because their output is not dynamically linked to the data used to generate it, violating the rule of automatic updating of the cells when any change is made in the spreadsheet. This leads to the danger that students can assume it is unnecessary (or simply forget) to re-run the macro when data are modified. Moreover, no trace is provided for which input cells were used to generate the outputs and simple inspection of one of these cells will reveal nothing. It is therefore a common feeling that the add-ins pose more problems than they solve (Hunt 1996; Nash and Quon 1996).

For a more advanced course the technical deficiencies of Excel are even more evident. Although many forms of graphics exist, most of them are designed not for statistical but for business users. Thus boxplots, stem and leafs diagrams, and q-q plots are not available. There are no non-parametric tests, principle components analysis or other multivariate analysis tools, multi-factor analysis of variance (e.g. Latin square), multiple comparisons procedures (e.g. Tukey test, Dunnet test), generalised linear models, survival analysis - all of which are standard features of a statistical package. Although Visual Basic has been used to extend the capabilities of the software (see for instance Jones, Hagtvedt and Jones, 2004), this approach will lead to an excessively programming-oriented course.

As a last point, there is a large literature that discuss the accuracy and reliability of statistical functions in Excel. As these papers point out, there is strong evidence that Excel can get very wrong answers for even simple statistical problems (see among others McCullough and Wilson 1999; Cryer 2001; McCullough 2001; Knüsel 2002; McCullough and Wilson 2001).

### 3. THE CHOICE OF A STATISTICAL PACKAGE

Several statistical packages are available on the market, among them SPSS (http://www.spss.com), Stata (http://www.stata.com), SAS (http://www.sas.com), Statistica (http://www.statsoft.com), S-plus (http://www.insightful.com), MINITAB

(http://www.minitab.com) and R (http://www.r-project.org). No one is an undisputed "killer application" and different conveniences and disadvantages exist for each.

The choice of a statistical package is usually institution-specific, motivated mainly by economics constraints (i.e. using the software available at the institution). This criterion can unfortunately lead to a far from optimal choice. For instance, when we had to choose among the above mentioned software packages, none of them was available in our informatics laboratory. Since we believe this decision as very relevant for statistics education, it seems appropriate to discuss in some detail the guidelines we adopted in making our choice:

a) Students must be able to access the software outside the course – i.e. they must be able to easily obtain a copy of the software to install on their personal computer. A prerequisite is thus a low cost for the license. This was identified as a key point. Students must be able to experiment themselves with the software, speeding up the learning process. It also produces long term benefits for the students attending the course. Since learning to use such software is a demanding task, it is convenient to introduce the students to a package that they can use in their professional life. This is not the case if the course is based upon very expensive software, because it is unlikely the institution that will employ them in the future will be inclined to invest money for software licences if they never used the software before.

b) Students must be able to learn to use the software in a reasonably short time. If the interaction with the statistical package is too complex there is the risk that most of the students will get discouraged early in the course. This will lead to a passive audience during the laboratory sessions, without any attempt to understand what is going on or to see general patterns.

c) Students must be able to access substantial documentation on the software. This implies that several tutorials are available, either published or on-line, and that the help system which comes with the software is good. The availability of such resources is an invaluable out-of-class reference for people attending the course and can allow the most willing students to expand, on their own, their knowledge of the software.

d) Students must be able to use the software for whatever analysis they need to do. This implies that the software must be as complete as possible, offering a large number of statistical tools, both graphical and computational. The software must be easily expansible by the user by the insertion of hand-written code; this permits treating problems that cannot be solved using the procedures supplied with the package.

e) As a last point, it is advisable that the chosen software is able to run under several different operating systems (at least Windows and GNU/Linux).

Several packages mentioned above have extensive capabilities for data manipulation and editing with a spreadsheet-like interface, a characteristic that can ease the transition of the students from Excel to these programs. Almost all of them have a rich graphical user interface (GUI), with the important exception of R, a command line software. The most common software among professional statisticians is probably S-plus, due to its power and ease of extending its capabilities to cope with unusual (non standard) problems. In effect, it relies on a programming language, known as S, and newly developed algorithms are presented in dedicated publications. Other users, if interested, can then easily adopt these algorithms without having to wait until a new release of the software is available from the developer. Although S-plus is widely used for teaching statistics at graduate levels, it has had less impact for mainstream undergraduate teaching despite radical approaches such as Nolan and Speed (1999, 2000).

The major drawback of adopting S-plus was the licence cost and the fact that the students were unable to access the package at a reasonable price.

The same was true for other packages such as SPSS, Stata and MINITAB, which are common among the researchers in Biology and Medicine. As a matter of fact the only statistical software considered which passes the constraint of the guideline (a) was R. It consists of a language plus a run-time environment with graphics, a debugger, access to certain system functions, and the ability to run programs stored in script files (R Development Core Team, 2004). The language is very similar in appearance to S, and most of the code written in S can run in R with no or little modification. R is free software distributed under a GNU-style copyleft, and an official part of the GNU project. All these features have made it very popular among academics.

A vast amount of background information, documentation, and other resources are available for R. CRAN (Comprehensive R Archive Network, http://www.r-project.org) is the centre for information about the R environment. The documentation includes an excellent internal help system, printed documentation (in particular "An Introduction to R"), a "frequently asked questions" list, and a number of textbooks (see the R project web page for a comprehensive bibliography). In addition to the textbooks, the R project web page hosts several contributed tutorials and manuals.

The main problem, however, is that the environment has a moderately steep learning curve due to the lack of a graphical interface (see guideline (b) mentioned above). Students need to learn the R syntax and remember the function names to use the program. It is therefore fundamental that teachers work out a soft approach to the R language introducing step by step the functions needed for statistical calculations. In summary, if R is certainly very powerful once mastered, it is non-trivial to learn and teach. In addition, data manipulation and editing are not as straightforward as they are when using a spreadsheet. As a consequence, all the phases preceding the statistical analysis take much more time than the analysis itself and are an error-prone procedure.

Our attempt to solve this problem was to integrate Excel, for data manipulation, and R, for statistical analysis. Although among statisticians it is common practice to use different programs in different phases of the analysis, it is uncommon to adopt a similar approach when planning a course. This is certainly due to the fact that the transfer of data from one format to another is often a troublesome task for novice students. Consequently, considerable attention was paid to developing a protocol for manipulating data within Excel, exporting them in a convenient format, and finally rereading them in R (see Section 4).

The use of command line software is a big challenge for teachers and students, but it has also important pedagogical advantages. In fact, it prevents students from playing with the graphical interface menus or searching for an analysis tool without exact knowledge of what the various statistical procedure are designed to do; this behavior frequently leads to a totally wrong analysis because of the selection of an unsuitable procedure. In contrast, the use of a language based statistical package helps students to develop their statistical thinking since they rely only on the knowledge grasped during the lectures. In addition, in a command line environment teachers need not cope with the problem that GUI based operations are difficult to discuss in written form, making it tricky to write some course notes.

Another point worth stressing is that since R is free software, teachers can inspect the source code to see exactly how the various procedures are implemented. This is extremely useful for avoiding possible confusion which can arise for students when a function is coded in a slightly non standard way. Using proprietary software, it is impossible to check what functions really do and warn students about anomalies and it is possible that when students solve textbook problems and get answers which differ from those reported in the text, their confidence is damaged.

## 4. THE BEST OF BOTH WORLDS. THE COMBINED USE OF EXCEL AND R

The interaction of Excel (Excel 2000) and R (R 1.8.1 and R 2.0.0) has been realized in a course of Biomedical Statistics for PhD students in Molecular Biology and Human Medicine. The duration of the course - held at Scuola Normale Superiore of Pisa (Italy) in 2003/04 and 2004/05 - was 28 hours, 12 of which were devoted to practical exercises in the informatics laboratory. The course is intended to make students able to analyze for themselves data collected during their professional activity. The knowledge of Statistics of graduate students in fields as Biology and Medicine is usually limited to Theory of Probability, Distributions, Sampling Theory and Test of Hypothesis, topics covered in introductory courses during the first University year. Only a very small fraction of them attended laboratory sessions during their previous statistics courses; in all these cases the software adopted was Excel. Only one student had acquired, in his research activity, some knowledge of a dedicated statistics package (Minitab).

During the laboratory sessions the students (on average about 15 per year) solve some statistical problems of interest in their field of work. They work under the teachers' supervision. A video projector helps the teachers illustrate the basic syntax of R and guide students during the most critical phases of the analysis.

In the first part of the course we discuss some problems to show how to solve them using both Excel and R. In this way we illustrate the different procedures that must be followed in the two environments. In the course, when facing more advanced problems, we emphasize that although non-specialized software like Excel is often sufficient to solve easy problems, it is not a good choice for more complex and complete investigations, for which a dedicated software package, such as R, is undoubtedly more appropriate. In our experience the use of Excel is, however, a deep-seated habit among the graduate students and they are inclined to use it to solve any sort of problem. This is particularly harmful and time-consuming when the power of the software is not sufficient to reach the pre-determined aim. We think this is a product of previous undergraduate courses, where too often overviews of the existence of statistical software with their basic potentialities are neglected. It would be desirable for students to face advanced problems from the beginning of their studies and to resort to dedicated software as the natural tool for statistical analysis.

The choice of using the two software packages simultaneously requires introducing in the first lessons both the syntax of R and a protocol that allows, in a simple way and without mistakes, the transfer data from Excel to R.

Since there is excellent help available within R, the initial effort has been to gently introduce some of the language's basic functions and operators that were evaluated as fundamental and were introduced from the beginning. They are related to the variable and vector assignment and manipulation:

- <-        The assignment operator
- c         This is a generic function which combines its arguments to form a vector
- factor    This function is used to encode a vector as a factor
- :         This operator generates regular sequences. E.g. 1:4 is the sequence 1, 2, 3, 4

- seq    Another operator that generates (more complex) regular sequences

The use of these basics was presented along the way during the analysis whenever the need of a particular function arose (mainly when learning to make the first graphs). We emphasize that a consistent part of the first practical lesson was dedicated to using the R built-in help. In fact, since the help pages have a common structure and are provided of a wide range of useful examples, they are an excellent didactical complement that is always available to the R users. Therefore it is worthwhile teaching students to use the help system from the very beginning hoping to make them able to learn autonomously. In light of our experience this approach makes the students able to master the language's basic functions and operators in less than one hour.

Later in the course, in particular when teaching ANOVA, it was helpful to introduce the function:
- tapply    Apply a function to each group of values given by a unique combination of the levels of certain factors

and talk a bit about statistical model definition in R, which requires some new operators:
- ~    This operator is used to separate the left- and right-hand sides in model formula
- *    In model formula context, this operator is used in the right-hand side of the formula when fitting a non additive (interaction) model
  E.g.  response ~ A * B   specifies a model including not only the factors A and B, but also their interaction

The R coding of statistical models can be very intimidating for a novice user (see for instance Chapter 11 of "An Introduction to R"). Aware of this, during the practical lessons we spent a good deal of time studying simple problems that require the use of additive models. Only when students were sufficiently confident with these did we present models with interaction.

The most difficult decision we had to face concerned the procedure for interfacing Excel and R. First, we had to choose a format that is supported by both software packages. Direct importation of Excel files in R was not an option since R is not (at present) able to read such a format. Furthermore we want to propose a protocol of exchange which can be extended easily to other spreadsheets. The most obvious choice is ASCII format but, during the test phase, we had several problems transferring data files containing multiple-word labels or missing data, which are frequent in Biology and Medicine datasets. These problems can be solved either by deep knowledge of R data reading functions (but not in all cases) or by modification of the spreadsheet data. The first option is really too tricky to rely on for a course and, from our experience, tends to seriously discourage the students. The second often needs too many modification to the data, substituting multiple words label with single word contractions and filling all missing values empty cells with a standard label. Furthermore, out of the classroom, this annoying work must be redone every time the dataset is expanded by other collaborators working on the original spreadsheet file.

Our preference was to choose the CSV (Comma Separated Value) format since it allows good management of fields containing either digits or strings (possibly with blanks). This choice solves the nasty problem of exact assignment of all the cell values in the spreadsheet to the correct field in an R data frame without having to play too much with R importation functions. Having chosen the preferred format of data exchange, we suggest the following standard procedure of saving CSV files from the used spreadsheet, consisting of four steps:
1. Data are manipulated within the spreadsheet to obtain the quantities needed during the analysis. In this phase values of various variables are screened for possible typos; variables that are calculated starting from other variables are evaluated. Students have to check that missing data are represented by empty cells.
2. All the relevant variables that must enter the analysis are arranged in a new file that contains only them (we recall that CSV format support only single-sheet files).
3. The new file is saved with CSV format. Depending on the configuration of the used spreadsheet, fields are separated by comma or by semicolon.
4. The file is read in R and the analysis can begin. We suggest the use of the function read.csv (or read.csv2, if data are separated by semicolon) specifying the option *na=""* for a correct reading of missing values, as in the following two examples:
Ex. 1 - read.csv(filename, head=TRUE, na="")   This call reads the file *filename*, treating empty fields as missing values (na=""), assuming that the first file row contains the name of the variables (head=TRUE). The function works correctly when commas separate the data;
Ex. 2 - read.csv2(filename, head=TRUE, na="", dec=",")   This is the correct call when data are separated by semicolons and commas are used as decimal separators.

This very simple protocol exploits the students' basic knowledge of Excel for all phases of data manipulation. After a few data transfer exercises students were able to do all the required steps very quickly. Its big advantage is that it makes unnecessary introducing most of the R functions for data manipulation, which represent a great challenge for novice users. In this sense the friendly Excel environment was used as GUI for the R statistical package.

The combined use of the two packages has proven useful to overcome the natural resistance of students to learn a completely new software tool. *The use of R is proposed not as a complete replacement of Excel but as an extension of its capabilities*. In our experience this approach enormously accelerates the students' learning of the R language because they can concentrate only on the functions that are strictly speaking related to data analysis.

## 5. CONCLUSIONS

The American Statistical Association curriculum guidelines, for example, for undergraduate programs in statistical science require familiarity with a standard software package and should encourage the study of data management (American Statistical Association, 2004). However, in undergraduate courses the use of such software is often neglected and spreadsheets – which are more familiar to students – are employed as standard tools. Since this software allows basic statistical analysis it is unfortunately common that students graduate without learning to use (or even knowing the existence of) dedicated statistical packages. Our experience fully demonstrates the need for software packages to teach introductory statistics courses. This is especially true for students who will usually face complex data analysis problems in their professional activity, such as students of Biology and Medicine.

We strongly suggest the use of R as a package for statistics teaching. This is mainly motivated by the free software license (GPL) of this software package. Every interested teacher can obtain a copy of the software and distribute it to the students without economics concerns. Furthermore, the availability of many extension libraries and the possibility to hand write ad-hoc code make this software package very appealing.  A broadly use of such a software for statistics education will allow an extremely useful knowledge sharing among teachers of statistics working in different areas and at different levels.

In this paper we have described how Excel and R can be combined for statistics education. A similar consideration can be extended to other spreadsheets, in particular to the one of the Open Office packages (Open Source software). This alternative leads to a full proprietary-software free course, and can be especially interesting for instructors working under operating system such as GNU/Linux or UNIX.

The integration of the two software packages and the protocol we have developed for PhD students is simple enough to be successfully adopted even in introductory courses. The use of a dedicated statistical package can help students familiarize themselves with it from the beginning of their studies and can lay the foundations for the future use of R as a tool for more complex statistical research.

## ACKNOWLEDGEMENTS

The authors wish to thank M.C. Prati who held the theoretical lectures of the course in Biomedical Statistics and Steve Shore for his reading of this manuscript and useful comments.